\documentstyle[11pt,newpasp,twoside,natbib,url]{article}
\def\edcomment#1{\iffalse\marginpar{\raggedright\sl#1\/}\else\relax\fi}
\def\iaucirc{{IAU~Circ.}}

\reversemarginpar

\begin{document}

\title{Recent Globular Cluster Searches with Arecibo and the Green
  Bank Telescope}

\author{S.~M.~Ransom$^{1,2}$, J.~W.~T.~Hessels$^1$, I.~H.~Stairs$^3$,
  V.~M.~Kaspi$^1$, D.~C.~Backer$^4$, L.~J.~Greenhill$^5$, \&
  D.~R.~Lorimer$^6$}

\affil{$^1$McGill University Physics Dept., 3600 University St.,
  Montreal, QC H3A~2T8, Canada} 
\affil{$^2$Center for Space Research,
  Massachusetts Institute of Technology, Cambridge, MA 02139}
\affil{$^3$Department of Physics and Astronomy, University of British
  Columbia, 6224 Agricultural Road, Vancouver, BC V6T~1Z1, Canada}
\affil{$^4$Dept. of Astronomy and Radio Astronomy Laboratory,
  University of California at Berkeley, 601 Campbell Hall 3411,
  Berkeley, CA 94720}
\affil{$^5$Harvard-Smithsonian Center for Astrophysics, 60 Garden St.,
  Cambridge, MA 02138}
\affil{$^6$University of Manchester, Jodrell
  Bank Observatory, Macclesfield, Cheshire SK11 9DL, UK}

\begin{abstract}
  We report the discovery of seven new millisecond pulsars during
  20\,cm searches of 19 globular clusters using the recently upgraded
  Arecibo Telescope and the new Green Bank Telescope (GBT).  Five of
  the pulsars are in compact binaries, three of which are eclipsing.
  One of the systems is in an eccentric orbit which is almost
  certainly highly relativistic.  Additional searches and timing
  observations are underway.
\end{abstract}

\section{Introduction}

In the past few years several groups have reported the discovery of 27
new millisecond pulsars (MSPs; 21 of which are in binary systems) in 8
globular clusters \citep{clf+00, pdm+01, ran01, jcb+02}, including an
amazing 11 new binary MSPs in 47~Tucanae alone (e.g. Lorimer et al.,
these proceedings).  The timing and multi-wavelength follow-up of
these systems has already resulted in a wealth of science on the
pulsar companions, the globular clusters, and the pulsars themselves
(e.g. \nocite{fcl+01} Freire et al., 2001; Heinke et al.~and Possenti
et al., these proceedings).

A significant fraction of the success of these searches can be
attributed to improvements in instrumentation and computational
capability.  At Parkes --- where most of these new pulsars have been
found --- the addition of a wide-bandwidth, low-temperature 20\,cm
receiver has revitalized cluster pulsar searching.  Earlier searches
were conducted almost exclusively at either 400 or 600\,MHz where
significant dispersion smearing, pulse scattering, and high galactic
background temperatures greatly reduce the system sensitivity.
Searching at 20\,cm mitigates these issues and in general improves
sensitivities even though the pulsars themselves are often fainter at
the higher frequency.  Additionally, the extensive (but
computationally expensive) use of acceleration searches on clusters
with known dispersion measures (DMs) has greatly improved
sensitivities to pulsars in compact binaries \citep[e.g.][]{jk91}.

With the availability of post-upgrade Arecibo and the GBT, a dedicated
104-processor computer cluster at McGill, and new binary pulsar search
algorithms \citep*{rem02, rce02}, we decided to search several of the
most promising northern globular clusters at 20\,cm.

\section{Observations and Data Analysis}

In 2001 June at Arecibo, we observed two full tracks (typically
1.5$-$2.5\,hrs) on each of 10 clusters (M3, M5, M13, M15, M53, M71,
NGC4147, NGC6426, NGC6760, and Pal2) using the Wideband Arecibo Pulsar
Processor (WAPP)\footnote{\url{http://www.naic.edu/~wapp}}.  Typical
observing parameters included 64\,$\mu$s sampling of 256 16\,bit lags
of 2 summed polarizations over a bandwidth of 100\,MHz and centered at
1475\,MHz.  With a data rate of $\sim$7.6\,MB/s, we accumulated
$\sim$1.5\,TB of search data.  An additional $\sim$1.5\,TB of search
data were taken in 2002 July on 12 other globular clusters and are
currently being processed.

In 2001 September and October at the GBT, we observed 12 clusters for
either 4\,hrs (M2, M4, M75, M80, M92, and NGC6342) or 8\,hrs (M3, M13,
M15, M30, M79, and Pal1) using one or two Berkeley-Caltech Pulsar
Machines\footnote{\url{http://www.gb.nrao.edu/~dbacker}}
\citep{bdz+97}.  Observations were made using 96$\times$1.4\,MHz
channels of 2 summed polarizations centered at 1375\,MHz and 4\,bit
sampled every 50\,$\mu$s.  We took a total of $\sim$0.5\,TB of data.
Significant quantities of both persistent and transient broadband
interference and the very strong Lynchburg airport radar with a period
of approximately 12\,s have made data analysis a challenge.
Approximately one quarter of these data remain to be fully analyzed.

We searched the data in several stages.  After attempting to remove
significant interference via time and frequency domain methods, we
de-dispersed the data over suitable ranges of DMs based either on the
average DM of the cluster for those with known pulsars or
$\sim$50$-$200\% of the predicted DM according to \citet{tc93}.  We
performed Fourier-domain acceleration searches and phase-modulation
(i.e. sideband) searches on the full observations and then additional
acceleration searches on overlapping short intervals of $\sim$15 and
$\sim$40\,min duration.  To date, we have discovered seven new MSPs
(see Table~1) and have several other good candidates awaiting
confirmation.


\begin{table}[t]
\centerline{{Table 1: Newly Discovered MSPs}}
\label{tab:psrs}
\begin{center}
\begin{tabular}{lcccccc}
\hline \hline
 & & $P_{psr}$ & DM & $P_{orbit}$ & $a_1 \sin (i)/c$       & Min $M_2^a$  \\
Pulsar & Telescope & (ms) & (pc cm$^{-3}$) & (hr) & (lt-s) & (M$_{\sun}$) \\
\hline \hline
M30A & GBT            & 11.0 & 25.1 & 4.18    & 0.23  & 0.10       \\
M30B & GBT            & 13.0 & 25.1 & $\ga$15 & Unk.  & $\sim$0.35 \\
M13C & GBT \& Arecibo & 3.72 & 30.1 &         &       &            \\
M13D & Arecibo        & 3.12 & 30.6 & 14.2    & 0.92  & 0.18       \\
M5C  & Arecibo        & 2.48 & 29.3 & 2.08    & 0.057 & 0.038      \\
M71A & Arecibo        & 4.89 & 117  & 4.24    & 0.078 & 0.032      \\
M3A  & Arecibo        & 2.54 & 26.5 & Unk.    & Unk.  & Unk.       \\
\hline \hline
\end{tabular}
\end{center}
$^a$ Assuming a pulsar mass ($M_1$) of 1.4\,M$_{\sun}$.
\end{table}

\section{The Pulsars}

\subsection{The Isolated MSP M13C}

M13C is a fairly typical isolated pulsar in M13 which was discovered
simultaneously in Arecibo and GBT observations.  Interstellar
scintillation causes significant changes (factors of 2$-$10) in the
measured flux density of M13C (and the three other known pulsars in
the cluster) over timescales of several hours.

\subsection{Three Eclipsing MSPs}

M30A, M5C, and M71A are all in very compact circular systems and
eclipse for $\sim$20\% of their orbital period at 20\,cm.  M30A and
M5C show evidence of time delays at both eclipse ingress and egress
(presumably due to dispersive delays as the pulses pass through the
ionized wind of the companion star), while M71A seems to simply slowly
fade and reappear at the eclipse ingress and egress.

M30A differs in several respects from the other two eclipsing MSPs.
It has a much longer pulse period (11\,ms as opposed to 2.5\,ms and
4.9\,ms for M5C and M71A respectively), it displays large changes in
flux density due to interstellar scintillation, and it likely has a
significantly more massive companion than the other two eclipsing
MSPs.  Archival {\em HST} imaging and recent {\em Chandra} data may
allow us to identify and characterize the companion star
\citep[e.g.][]{egh+01}.  Optical and/or X-ray follow-up on the nature
of the companions to M5C and M71A may also be possible once timing
positions are available.

\subsection{Other Binaries}

M3A has only recently been discovered and confirmed in Arecibo data
taken this past summer and hence the orbital solution is currently
unknown.  However, the large accelerations measured in the searches
imply that it is most likely in a compact binary.  Since M3A
scintillates significantly, we anticipate discovering more MSPs in
this cluster.

M13D is in a longer period binary ($\sim$14\,hrs) with a near circular
orbit ($e \sim 0.0006$) around a more massive companion
($\ga$0.18\,M$_{\sun}$).  A several year timing baseline should allow
us to measure the rate of change of the angle of periastron,
$\dot{\omega}$, and hence determine the total mass of the system.

M30B is perhaps the most intriguing of the new MSPs.  It was
discovered during the initial observation of M30 at the same DM as
M30A.  The measured significances were $\sim$30\,$\sigma$ and
$\sim$10\,$\sigma$ from the two BCPMs, which had different center
frequencies, total bandwidths, and interference conditions.  But in
the more than 20 GBT observations of M30 taken since then, we have not
seen it again.  In the $\sim$7\,hrs in which it was visible, however,
we measure {\em four} significant frequency derivatives of the pulsar
due to the orbital motion.  Using these derivatives and the technique
described by \citet{jr97}, we determine that the orbit must have $e
\ga 0.35$, an orbital period of $\sim$1$-$10\,days, and a companion
mass $\ga$0.35\,M$_{\sun}$, making it a highly relativistic system
with the potential to accurately measure the component masses and test
gravitational theories.

\section{Future Prospects}

We are timing each of the new MSPs mentioned above and analyzing the
remaining search data from Arecibo and the GBT.  In the near future,
better pulsar backends at Arecibo (4 WAPPs covering 400\,MHz of
bandwidth) and the GBT (the $\sim$600\,MHz bandwidth ``Spigot'' card)
and a much improved 20\,cm receiver at Arecibo will improve our timing
precision and allow us to find fainter pulsars in the timing data.
Observations at other promising frequencies (i.e.  375\,MHz at the
GMRT and 800\,MHz at the GBT) may uncover even more pulsars in these
clusters.  Finally, analysis of optical and X-ray observations will
help to maximize the science output from these remarkable systems.


\end{document}